\documentclass[twocolumn]{IEEEtran}
                                                                                
\usepackage{amsmath,amstext,amssymb,epsf}

 \newtheorem{lemma}{Lemma}[section]
 \newtheorem{theorem}[lemma]{Theorem}

 \newtheorem{claim}[lemma]{Claim}

 \newtheorem{definition}[lemma]{Definition}
\newtheorem{rem}[lemma]{Remark}
\newenvironment{remark}{\begin{rem}}{\hspace*{\fill}$\diamondsuit$\end{rem}}
 \newtheorem{ex}[lemma]{Example}
\newenvironment{example}{\begin{ex}}{\hspace*{\fill}$\diamondsuit$\end{ex}}
                                                                                
\numberwithin{equation}{section} 

\usepackage{psfig}

\begin{document}
\title{The Similarity Metric}
\author{Ming Li, Xin Chen, Xin Li, Bin Ma, and Paul M.B. Vit\'anyi
\thanks{The material of this paper was presented in part in
{\em Proc. 14th ACM-SIAM Symposium on Discrete Algorithms},
2003, pp 863-872.
Ming Li is with the
Computer Science Department, University of Waterloo, Waterloo,
Ontario N2L 3G1, Canada, 
and with BioInformatics Solutions Inc.,
Waterloo, Canada. He is partially supported by NSF-ITR grant 0085801
and NSERC.
Email: {\tt mli@wh.math.uwaterloo.ca}.
Xin Chen is with the Department of Computer Science, University
of California, Santa Barbara, CA 93106, USA. 
Email: {\tt chxin@cs.ucsb.edu}.
Xin Li is with the Computer Science Department, University
of Western Ontario, London, Ontario N6A 5B7, Canada. Partially
supported by NSERC grant RGP0238748. Email: {\tt xinli@csd.uwo.ca}.
Bin Ma is with the
Computer Science Department, University
of Western Ontario, London, Ontario N6A 5B7, Canada. Partially
supported by NSERC grant RGP0238748. Email: {\tt bma@csd.uwo.ca}.
Paul Vit\'anyi
is with the 
CWI, Kruislaan 413,
1098 SJ Amsterdam, The Netherlands, and with the University
of Amsterdam, Amsterdam, The Netherlands.
Email {\tt Paul.Vitanyi@cwi.nl}.
Partially supported by the
EU project QAIP, IST--1999--11234, the EU Project RESQ, 
the NoE QUIPROCONE IST--1999--29064,
the ESF QiT Programmme, the EU 
 NeuroCOLT II Working Group
EP 27150, and the EU PASCAL NoE.
}}

\markboth{IEEE Transactions on Information Theory, VOL. XX, NO Y, MONTH 2004}{Chen, Li, Li, Ma, Vit\'anyi: The Similarity Metric}

\maketitle

\begin{abstract}
A new class of distances
appropriate for measuring 
similarity relations between sequences, say one type of similarity
per distance, is studied.
We propose a new ``normalized information distance'',
based on the noncomputable notion of Kolmogorov complexity, and
show that it is in this class and it minorizes every computable distance
in the class (that is, it is universal in that
it discovers all computable similarities).
We demonstrate that it is a metric and
call it  the {\em similarity metric}.
This theory forms the foundation for a new practical tool.
To evidence generality and robustness
we give two distinctive applications in widely divergent areas
using standard compression programs like gzip and GenCompress.
First, we 
compare whole mitochondrial genomes and infer
their evolutionary history. This results in a first 
completely automatic computed
whole mitochondrial phylogeny tree. 
Secondly, we fully automatically compute the
language tree of 52 different languages.


\end{abstract}

\section{Introduction}

How do we measure similarity---for example to determine 
an evolutionary distance---between two sequences,
such as internet documents, different language text corpora in the same
language, among different languages based on example text corpora,
computer programs, or chain letters?
How do we detect plagiarism of student source code in assignments? 
Finally, the fast advance of worldwide genome sequencing 
projects has raised the following
fundamental question to prominence in contemporary biological science: how
do we compare two genomes
\cite{Koonin1999,Wooley1999}? 

Our aim here is not to define a similarity measure for a certain
application field based on background knowledge
and feature parameters specific to that field; instead
we develop a general
mathematical theory of similarity that uses no background knowledge
or features specific to an application area.  Hence it is,
without changes, applicable to different areas and even 
to collections of objects taken from different areas.
The method automatically zooms in on the dominant similarity aspect
between every two objects. To realize this goal, we first define 
a wide class of similarity distances. Then, we
show that this class contains a particular distance
that is universal in the following sense: for every pair of objects
the particular distance is less than any ``effective''
distance in the class 
between those two objects.   
This universal distance is called the ``normalized information distance'' (NID),
it is shown to be a metric, and, intuitively, it 
uncovers all similarities simultaneously that effective distances in the class
uncover a single similarity apiece. 
(Here, ``effective'' is used as shorthand 
for a certain notion of ``computability''
that will acquire its precise meaning below.) 
We develop a practical analogue of the NID
based on real-world compressors, called the ``normalized compression
distance'' (NCD), and test it
on real-world applications in a wide range of fields:
we present
the first completely automatic construction
of the phylogeny tree based on whole mitochondrial genomes, and
a completely automatic construction of a language tree for over 50
Euro-Asian languages. 

{\bf Previous Work:}
Preliminary applications of the current approach were
tentatively reported to the biological community and
elsewhere \cite{CKL99,LBCKKZ01,LV01}.
That work, and the present paper, is based on {\em information distance}
\cite{LiVi97,BGLVZ98}, a universal metric that minorizes in
an appropriate sense every
effective metric: effective versions of Hamming distance, 
Euclidean distance, edit distances,
Lempel-Ziv distance, and the sophisticated distances introduced
in \cite{CPSV00,MS00}.
Subsequent work in the
linguistics setting, \cite{BCL02a,BCL02b}, 
used related ad hoc
compression-based methods, 
Appendix~\ref{app.A}. 
The information distance studied in \cite{LiVi96,LiVi97,BGLVZ98,LBCKKZ01},
and subsequently investigated in \cite{hammer97,MV01,SV00,VV00},
is defined as 
the length of the shortest binary program 
that is needed to transform the two objects into each other.
This distance can be interpreted also
as being proportional to the minimal amount of 
energy required to do the transformation:
A species may lose genes (by deletion) or gain genes (by 
duplication or insertion from external sources), relatively 
easily. Deletion and insertion cost energy
(proportional to the Kolmogorov complexity of deleting or inserting
sequences in the information distance), and aspect that
was stressed in \cite{LiVi96}.
But this distance is not proper to measure evolutionary 
sequence distance. For example, {\em H. influenza} and {\em E. coli}
are two closely related sister species. The former has about 1,856,000
base pairs and the latter has about 4,772,000 base pairs. 
However, using the information distance of \cite{BGLVZ98}, one
would easily classify {\em H. influenza} with a short (of comparable
length) but irrelevant species simply because of length, instead of 
with {\em E. coli}. The problem is that the information distance
of \cite{BGLVZ98} deals with {\em absolute} distance
rather than with {\em relative} distance.
The paper \cite{Varre1998} defined a transformation distance between two
species, and \cite{Grumbach1994} defined a compression distance. 
Both of these measures are essentially related to $K(x|y)$. Other than being
asymmetric, they also suffer from being absolute rather than relative.
As far as the
authors know, the idea of relative or normalized distance is, surprisingly, not
well studied. An exception is \cite{Ya02}, which investigates
normalized Euclidean metric and
normalized symmetric-set-difference metric to account for relative
distances rather than absolute ones, and it does so
for much the same reasons as does the
present work. In \cite{Ra61} the equivalent functional of \eqref{distance.s}
in information theory, expressed in terms of  
the corresponding probabilistic notions, is shown to
be a metric. (Our Lemma~\ref{lem.metric} implies this result, 
but obviously not the other way around.)

{\bf This Work:}
We develop 
a general
mathematical theory of similarity based on a notion
of normalized distances. 
Suppose we define a new distance by setting the value
between every pair of objects to the minimal upper semi-computable
(Definition~\ref{def.comp} below)
normalized distance (possibly a different distance for every pair). 
This new distance is a non-uniform lower bound on the upper semi-computable
normalized distances.
The central notion of this work is the ``normalized information distance,''
given by a simple formula,
that is a metric, 
belongs to the class of normalized distances, and minorizes
the non-uniform lower bound above. 
It is (possibly) not upper semi-computable, but it
is the first {\em universal} 
similarity
measure, and is an objective recursively invariant notion by the 
Church-Turing thesis \cite{LiVi97}. 
We cannot compute the normalized information distance,
which is expressed in terms of the noncomputable Kolmogorov
complexities of the objects concerned. Instead, we look at wether
a real-world imperfect analogue works experimentally, by replacing
the Kolmogorov complexities by the length of the compressed objects
using real-world compressors like gzip or GenCompress.  
Here we show the results of experiments in
the diverse
areas of (i) bio-molecular evolution studies,
 and (ii) natural language evolution. 
In area (i): In recent
years, as the complete genomes of various species become available,
it has become possible to do whole genome phylogeny (this overcomes the 
problem that different genes may give different trees \cite{Cao1998,RWKC03}). 
However, 
traditional phylogenetic methods on individual genes depended
on multiple alignment of the related proteins
and on the model of evolution of individual amino acids.
Neither of these is practically applicable to the genome level.
In this situation, a method that can compute shared
information between two individual sequences is useful because biological
sequences encode information, and the occurrence of evolutionary
events (such as insertions, deletions, point mutations,
rearrangements, and inversions) separating two sequences sharing a
common ancestor will result in partial loss of their shared
information. Our theoretical approach is used experimentally to
create a fully automated and reasonably accurate
software tool based on such a distance to compare two
genomes. We demonstrate that
a whole mitochondrial genome phylogeny of the Eutherians 
can be reconstructed automatically from {\em
unaligned} complete mitochondrial genomes by use of our software 
implementing (an approximation of) our theory, 
confirming one of the hypotheses in \cite{Cao1998}.
These experimental confirmations of the effacity
of our comprehensive approach contrasts
with recent more specialized approaches such 
as \cite{WW01} that have (and perhaps
can) only be tested on
small numbers of genes. They have not been experimentally tried on
whole mitochondrial genomes that are, apparently,
already numerically out of computational range.
In area (ii) we fully automatically 
construct the language tree of 52 primarily
Indo-European languages from translations of the 
``Universal Declaration of Human Rights''---leading to a
grouping of language families 
largely consistent with current linguistic viewpoints. 
Other experiments and applications performed earlier,
not reported here are: detecting
plagiarism in student programming assignments \cite{SID}, phylogeny
of chain letters in \cite{BLM03}.

{\bf Subsequent Work:}
The current paper can be viewed as the theoretical basis out of
a trilogy of papers:
In \cite{CV03} we address the gap between the rigorously
proven optimality of the normalized information distance based
on the noncomputable notion of Kolmogorov complexity, and the
experimental successes of the ``normalized compression distance''
or ``NCD''
which is the same formula with the Kolmogorov complexity
replaced by the lengths in bits of the compressed files
using a standard compressor.
We provide an axiomatization of a notion of ``normal compressor,'' 
and argue that
all standard compressors, be it of
the Lempel-Ziv type (gzip), block sorting type (bzip2), or statistical type
(PPMZ), are normal.
It is shown that the NCD based on a normal compressor
is a similarity distance, satisfies the metric properties, and
it approximates universality.
To extract a hierarchy of clusters
from the distance matrix,
we designed 
a new quartet
method and a fast heuristic to implement it.
The method is implemented and 
available on the web as a free open-source software tool: the CompLearn
Toolkit \cite{Ci03}.
To substantiate claims of universality and robustness,
\cite{CV03} reports successful applications in areas as diverse as
genomics, virology, languages, literature, music, handwritten digits,
astronomy, and
combinations of objects from completely different
domains, using statistical, dictionary, and block sorting compressors.
We tested the method both on natural data sets 
from a single domain and combinations of different domains
(music, genomes, texts, executables, Java programs),
and on artificial ones where we know the right answer. 
In \cite{CVW03} we applied the method in detail to to music clustering,
(independently \cite{LLB03} applied the method of \cite{BCL02a} in
this area). The method has been reported abundantly and extensively in 
the popular science press, for example \cite{Mu03,Pa03,BLM03,De04},
and has created considerable attention, and follow-up applications
by researchers in
specialized areas. 
One example of this is in parameter-free data mining 
and time series analysis \cite{KLR04}. In that paper the effacity
of the compression
method is evidenced by a host of experiments. It is also shown that
the compression based method,
is superior to any other method for comparision
of heterogeneous files (for example time series),
and anomaly detection,
see Appendix~\ref{app.B}, 

\section{Preliminaries}
\label{sect.prel}
{\bf Distance and Metric:}
Without loss of
generality, a distance only needs to operate on finite sequences of 0's and
1's since every finite sequence over a finite alphabet
can be represented by a finite binary sequence.
Formally, a {\em distance} is a function $D$ with nonnegative
real values, defined on the Cartesian product $X \times X$ of
a set $X$. It is called a {\em metric}
on $X$ if for every $x,y,z \in X$:
\begin{itemize}
\item
$D(x,y)=0$ iff $x=y$ (the identity axiom);
\item
$D(x,y)+D(y,z) \geq D(x,z)$ (the triangle inequality);
\item
$D(x,y)=D(y,x)$ (the symmetry axiom).
\end{itemize}
A set $X$ provided with a metric is called a {\em metric space}.
For example, every set $X$ has the trivial {\em discrete metric}
$D(x,y)=0$ if $x=y$ and $D(x,y)=1$ otherwise.

{\bf Kolmogorov Complexity:}
A treatment of the theory of Kolmogorov
complexity can be found in the text \cite{LiVi97}. Here we recall some
basic notation and facts.
  We write {\em string} to mean a finite binary string.
  Other finite objects can be encoded into strings in natural
ways.
  The set of strings is denoted by $\{0,1\}^*$. 
The Kolmogorov complexity of a file
is essentially the length of the ultimate compressed version of the file.
  Formally, the {\em Kolmogorov complexity}, 
or {\em algorithmic entropy}, $K(x)$ of a
string $x$ is the length of a shortest binary program $x^*$ to compute
$x$ on an appropriate universal computer---such as a universal Turing machine.
Thus, $K(x)=|x|$, the {\em length} of $x^*$ \cite{Ko65}, denotes the
number of bits of information from which $x$ can be computationally
retrieved. If there are more
than one shortest programs, then $x^*$ is the first one in standard
enumeration.
\begin{remark}
\rm
We require that there $x$ can be decompressed from its
compressed version $x^*$ by a general decompressor program, 
but we do not require
that $x$ can be compressed to $x^*$ by a general compressor program.
In fact, it is easy to prove that there does not exist such
a compressor program, since $K(x)$ is a noncomputable function.
Thus, $K(x)$ serves as the ultimate, lower bound  
of what a real-world compressor can possibly achieve.
\end{remark}
\begin{remark}
\rm
To be precise, without going in details,
the Kolmogorov complexity we use is the ``prefix'' version, where
the programs of the universal computer are prefix-free (no program
is a proper prefix of another program).  
It is equivalent to consider the length of 
the shortest binary program to compute
$x$ in a universal programming language such as LISP or Java.
Note that these programs are always prefix-free, since there is
an end-of-program marker.
\end{remark}
  The {\em conditional} Kolmogorov complexity $K(x \mid y)$ of $x$ relative to
$y$ is defined similarly as the length of a shortest program
to compute $x$ if $y$ is furnished as an auxiliary input to the
computation. We use the notation $K(x,y)$ for the length of a
shortest binary program that prints out $x$ and $y$ and a description
how to tell them apart. 
  The functions $K( \cdot)$ and $K( \cdot| \cdot)$,
though defined in terms of a
particular machine model, are machine-independent up to an additive
constant
 and acquire an asymptotically universal and absolute character
through Church's thesis, from the ability of universal machines to
simulate one another and execute any effective process.

\begin{definition}\label{def.comp}
A real-valued function $f(x,y)$ is {\em upper semi-computable} if there exists
a rational-valued recursive function $g(x,y,t)$ such that (i)  $g(x,y,t+1) \leq g(x,y,t)$, and
(ii) $\lim_{t \rightarrow \infty} g(x,y,t)=f(x,y)$. It is {\em lower semi-computable}
if $-f(x,y)$ is upper semi-computable, and it is {\em computable} if 
it is both upper- and lower semi-computable. 
\end{definition}

It is easy to see that the 
functions $K(x)$ and $K(y \mid x^*)$ (and under the appropriate interpretation
also $x^*$, given $x$) are 
upper semi-computable, and it is easy to prove that they are not computable. 
The conditional information contained in $x^*$ 
is equivalent to that in $(x,K(x))$: there are fixed
recursive functions $f,g$ such that for every $x$ we have
$f(x^*)=(x,K(x))$ and $g(x,K(x))=x^*$. 
The {\em information about $x$ contained in $y$} is defined as 
$I(y:x)=K(x)-K(x \mid y^*)$. A deep, and very useful, result 
\cite{Ga74} shows that 
there is a constant $c_1 \geq 0$, independent of $x,y$, such that
\begin{equation}\label{eq.soi}
K(x,y)=K(x)+K(y \mid x^*) = K(y)+K(x \mid y^*),
\end{equation}
with the equalities holding up to $c_1$ additive precision.
Hence, up to an additive constant term $I(x:y) = I(y:x)$.

{\bf Precision:}
It is customary in this area to use ``additive constant $c$'' or
equivalently ``additive $O(1)$ term'' to mean a constant,
accounting for the length of a fixed binary program,
independent from every variable or parameter in the expression
in which it occurs.

\section{Information Distance}
In our search for the proper definition of the distance between
two, not necessarily equal length, binary strings, a natural choice
is the length of the shortest program that can transform either 
string into the other one---both ways, \cite{BGLVZ98}. This 
is one of the main concepts
in this work. Formally,
  the {\em information distance} 
is the length $E(x,y)$ of a shortest binary program that
computes $x$ from $y$ as well as computing $y$ from $x$.
  Being shortest, such a program should take advantage of any
redundancy between the information required to go from $x$ to $y$
and the information required to go from $y$ to $x$.
  The program functions in a catalytic capacity in the
sense that it is required to transform the
input into the output, but itself remains
present and unchanged throughout the computation.
 A principal result of \cite{BGLVZ98} shows that
the information distance equals
 \begin{equation}\label{eq.infodistance}
E (x,y)=  \max\{K(y \mid x),K(x \mid y)\} 
 \end{equation}
up to an additive $O(\log  \max\{K(y \mid x),K(x \mid y)\})$ term.
The information distance $E(x,y)$ is upper
semi-computable:
By dovetailing the running of all programs we can find
shorter and shorter candidate prefix-free programs $p$ with 
$p(x)=y$ and $p(y)=x$, and in the limit obtain such a $p$
with $|p|=E(x,y)$. 
  (It is very important here that the time of computation is
completely ignored: this is why this result does not contradict the
existence of one-way functions.) 
It was shown in \cite{BGLVZ98}, Theorem 4.2, that
the information distance $E(x,y)$ is a metric. More precisely,
it satisfies the metric
properties up to an additive fixed finite constant.
A property of $E(x,y)$ that is central for our purposes here is that
it minorizes every ``admissible distance'' (below) up to an additive constant.
In defining the class of admissible distances
we want to exclude unrealistic distances
like $f(x,y) = \frac{1}{2}$ for {\em every} pair $x  \neq y$, by restricting
the number of objects within a given distance of an object. Moreover,
we want distances to be computable in some manner.
\begin{definition}\label{def.em}
\rm
Let $\Omega = \{0,1\}^*$.
A function $D: \Omega \times \Omega \rightarrow {\cal R}^+$ (where ${\cal R}^+$
denotes the positive real numbers) is an
{\em admissible distance} if it is upper semi-computable, symmetric,
and for every pair of objects $x,y \in \Omega$
the distance  $D(x,y)$ is
the length of a binary prefix code-word that is a program that computes
$x$ from $y$, and vice versa, in the reference programming language.
\end{definition}
\begin{remark}
\rm
In \cite{BGLVZ98} we considered ``admissible metric'',
but the triangle inequality metric restriction is not necesary for our purposes
here.
\end{remark}
If $D$ is  an admissible distance, then for every $x\in \{0,1\}^*$ 
the set $\{D(x,y): y \in \{0,1\}^*\}$ is the length set of a prefix code.
Hence it satisfies the Kraft inequality \cite{LiVi97}, 
\begin{equation}\label{eq.dc}
\sum_{y} 2^{-D(x,y)} \leq 1,
\end{equation}
which gives us the desired density condition.

\begin{example}\label{ex.ham}
\rm
In representing the Hamming distance $d$ between
$x$ and $y$ strings of equal length $n$ differing in positions
$i_1, \ldots , i_d$, we can use a simple
prefix-free encoding of $(n,d,i_1, \ldots , i_d)$
in $H_n(x,y)=2 \log n + 4 \log \log n +2 + d \log n$ bits.
We encode $n$ and $d$ prefix-free in $\log n+2 \log \log n +1$
bits each, see e.g. \cite{LiVi97},
and then the literal indexes of the actual flipped-bit
positions. Hence, $H_n(x,y)$ is the length of a prefix code word (prefix
program) to compute $x$ from $y$ and {\em vice versa}.
Then, by the Kraft inequality,
\begin{equation}\label{eq.dch}
\sum_{y} 2^{-H_n(x,y)} \leq 1.
\end{equation}
It is easy to verify that $H_n$ is a metric in the sense that it satisfies
the metric (in)equalities up to $O(\log n)$
additive precision.
\end{example}

\begin{theorem}\label{th.id}
The information distance $E(x,y)$ is
an admissible distance that satisfies the metric inequalities
up to an additive constant,
and it is minimal in the sense that for every admissible distance
$D(x,y)$ we have
 \[
  E(x,y)\leq D(x,y)+O(1).
 \]
\end{theorem}
\begin{remark}
\rm
This is the same statement as Theorem 4.2 in \cite{BGLVZ98}, except that
there the $D(x,y)$'s were also required to be metrics. But the proof
given doesn't use that restriction and therefore suffices for
the slightly more general theorem as stated here. 
\end{remark}

Suppose we want to quantify how much objects differ in terms of
a given feature, for example the length in bits of files,
the number of beats per second in music pieces, the number
of occurrences of a given base in the genomes. Every specific
feature induces a distance, and every specific distance measure
can be viewed as a quantification of an associated feature difference.
The above theorem states that among all features that correspond to
upper semi-computable distances, that satisfy the density 
condition \eqref{eq.dc}, the information distance is universal in
that among all such distances it is always smallest up to constant
precision. That is, it accounts for the dominant feature in which two
objects are alike. 

\section{Normalized Distance}
Many distances
are absolute, but if we want to express similarity,
then we are more interested in relative ones.
For example, if two strings of length $10^6$ differ by 1000 bits,
then we are inclined to think that those strings are relatively 
more similar than two strings of 1000 bits that have that distance and
\begin{definition}\label{eq.defsm}\label{def.nm}
\rm
A {\em normalized distance} or {\em similarity distance},
is a function $d: \Omega \times \Omega \rightarrow [0,1]$
that is symmetric $d(x,y)=d(y,x)$, and 
for every $x \in \{0,1\}^*$ and every constant $e \in [0,1]$ 
\begin{equation}\label{eq.dp}
 | \{y: d(x,y) \leq e \leq 1 \} | < 2^{eK(x)  +1} .
\end{equation}
\end{definition}

The {\em density} requirement \eqref{eq.dp} 
is implied by a ``normalized'' version
of the Kraft inequality:
\begin{lemma}\label{lem.ki}
Let $d: \Omega \times \Omega \rightarrow [0,1]$ satisfy
\begin{equation}\label{eq.nc}
\sum_{y} 2^{- d(x,y)K(x)} \leq 1 .
\end{equation}
Then,  $d$  satisfies \eqref{eq.dp}.
\end{lemma}
\begin{proof}
For suppose the contrary:
there is an $e \in [0,1]$, such that
\eqref{eq.dp} is false. Then,
starting from (\ref{eq.nc}) we obtain
a contradiction:
\begin{align*}
1 & \geq \sum_{y} 2^{-d(x,y) K(x)}
\\ &\geq  \sum_{y : d(x,y) \leq e \leq 1}
2^{-eK(x)}
\\&  \geq 2^{eK(x)  +1} 2^{-eK(x)} > 1.
\end{align*}
\end{proof}

\begin{remark}
\rm
If $d(x,y)$ is a normalized version of an admissible distance $D(x,y)$
with $D(x,y)/d(x,y) \geq K(x)$, then \eqref{eq.nc} implies \eqref{eq.dc}.
\end{remark}

We call a normalized distance a ``similarity'' distance, because
it gives a relative similarity
(with distance 0 when objects are maximally similar and distance 1 when
the are maximally dissimilar)
and, conversely, for a well-defined notion of absolute distance
(based on some feature) we can express similarity according to
that feature 
as a similarity distance being a normalized version
of the original absolute distance.
In the literature a distance that expresses lack of similarity (like
ours)  is often
called a ``dissimilarity'' distance or a ``disparity'' distance.

\begin{example}
\rm
The prefix-code for the Hamming distance $H_n(x,y)$ between
$x,y \in \{0,1\}^n$ in
Example~\ref{ex.ham} is a program to compute from $x$ to $y$ and
{\em vice versa}. 
To turn it into a similarity distance 
define $h_n(x,y) = H_n(x,y)/(\alpha (x,y) n \log n)$ with
$\alpha (x,y)$ satisfying the inequality $n H( e \alpha (x,y)) \leq eK(x)$
for every $0 \leq e \leq 1$ and
$0 \leq h(x,y) \leq 1$ for every $n,x,y$,
where this time $H$ denotes the 
entropy with two possibilities with probabilities $p= en(x,y)$
and $1-p$, respectively.  For example, for 
$x$ with $K(x)=n$ and $y$ is within $n/2$ bit flips
of $x$, we can set $\alpha (x,y) = \frac{1}{2}$, yielding
$h_n(x,y) = 2d/n$ with $d$ the number of bit flips to obtain $y$
from $x$. 
For every $x$, the number of $y$ in the Hamming ball $h_n(x,y) \leq e$
is upper bounded by $2^{nH(e \alpha (x,y))}$. 
By the constraint on $\alpha(x,y)$, the  
function $h_n(x,y)$ satisfies the density condition
\eqref{eq.dp}.
\end{example}

\section{Normalized Information Distance}
Clearly, unnormalized information distance (\ref{eq.infodistance}) 
is not a proper evolutionary
distance measure. Consider three species: {\em E. coli}, {\em H. influenza},
and some arbitrary bacteria $X$ of similar length as {\em H. influenza},
but not related. Information distance $d$ 
would have $d(X, H. influenza) < d(E.coli, H. influenza)$,
simply because of the length factor. It would put
two long and complex sequences that differ only by a tiny fraction
of the total information as dissimilar as two short sequences that
differ by the same absolute amount and are completely random with respect
to one another. 
In \cite{LBCKKZ01} we considered as first attempt at a normalized 
information distance: 

\begin{definition} 
Given two sequences $x$ and $y$, define the function
$d_s(x,y)$ by
\begin{equation}\label{distance.s}
d_s(x,y) = \frac{K(x \mid y^*)+K(y \mid x^*)}{K(x,y)}.
\end{equation}
\end{definition}
Writing it differently, using \eqref{eq.soi}, 
\begin{equation}\label{eq.ds}
d_s(x,y) = 1- \frac{I(x:y)}{K(x,y)},
\end{equation}
where $I(x:y)=K(y)-K(y \mid x^*)$ is known as the {\em mutual 
algorithmic information}. It is ``mutual''
since we saw from \eqref{eq.soi} that it is symmetric: $I(x:y)=I(y:x)$
up to a fixed additive constant.
This distance satisfies the triangle inequality, up to a
small error term, and universality
(below), but only within a factor 2. 
Mathematically more precise and satisfying is the distance:

\begin{definition}
Given two sequences $x$ and $y$, define the function
$d(x,y)$ by
\begin{equation}\label{distance.m}
d(x,y) = \frac{\max\{K(x \mid y^*),K(y \mid x^*)\}}{\max \{K(x),K(y)\}}.
\end{equation}
\end{definition}

\begin{remark}
\rm
Several natural alternatives for the denominator
turn out to be wrong:

(a) 
Divide by the length. Then, 
firstly we do not know which of the two length involved
to divide by, possibly the sum or maximum,  but furthermore
the triangle inequality and the universality (domination)
properties are not satisfied. 

(b) In the $d$ definition divide by $K(x,y)$. Then one
has $d(x,y)= \frac{1}{2}$ whenever $x$ and $y$ are random (have maximal Kolmogorov
complexity) relative to one another. This is improper.

(c) In the $d_s$ definition dividing by length does not satisfy 
the triangle inequality.
\end{remark}

There is a natural interpretation to $d(x,y)$: If $K(y) \geq K(x)$
then we can rewrite
\[d(x,y) = \frac{K(y)-I(x:y)}{K(y)} = 1 - \frac{I(x:y)}{K(y)} . \]
That is, $1-d(x,y)$ between $x$ and $y$ is the
number of bits of information that is shared between the two strings
per bit of information of the string with most information.

\begin{lemma}\label{lem.metric}
$d(x,y)$ satisfies the metric (in)equalities up to additive 
precision $O(1/K)$, where $K$ is the maximum of the Kolmogorov
complexities of the objects involved in the (in)equality.
\end{lemma}
\begin{proof}
Clearly, $d(x,y)$ is precisely symmetrical. It
also satisfies the identity axiom up to the required precision: 
$$
 d(x,x)=O(1/K(x)). 
$$
To show that it is
a metric up to the required precision,
it remains to prove the triangle inequality.

\begin{claim}
$d(x,y)$ satisfies the triangle inequality $d(x,y) \leq
d(x,z) + d(z,y)$ up to an additive error term of $O(1/\max\{K(x),K(y),K(z)\})$.
\end{claim}

\begin{proof}
{\bf Case 1:} Suppose $K(z) \leq \max\{K(x),K(y)\}$.
In \cite{GTV01}, the following ``directed triangle inequality''
was proved:
For all $x,y,z$, up to an additive constant term,
 \begin{equation}\label{eq.magic}
  \hspace{0.7cm}K(x \mid y^*) \leq K(x, z \mid y^{*}) \leq  K(x \mid z^*)+ K(z \mid y^*).
\end{equation}



\noindent
Dividing both sides by
$\max\{K(x),K(y)\}$, majorizing and rearranging,
\begin{eqnarray*}
  &&  \frac{\max\{K(x \mid y^*) ,K(y \mid x^*)\}}{\max\{K(x),K(y)\}} 
 \\&& \hspace{-0.9cm} =
\frac{\max\{K(x \mid z^*)+K(z \mid y^*), K(y \mid z^*)+K(z \mid x^*)\}}
{\max\{K(x),K(y)\}}
 \\&& \hspace{-0.9cm} \leq 
 \frac{\max\{K(x \mid z^*),K(z \mid x^*) \}}{\max\{K(x),K(y)\}}
+
 \frac{ \max\{K(z \mid y^*),K(y \mid z^*)\}}{\max\{K(x),K(y)\}},
\end{eqnarray*}
up to an additive term $O(1/\max\{K(x),K(y),K(z)\})$.
Replacing $K(y)$ by $K(z)$ in the denominator of the first
term in the right-hand side, and  $K(x)$ by $K(z)$ in the
denominator of second term of the right-hand side, respectively,
can only increase the right-hand side (again, because of the assumption).

{\bf Case 2:} Suppose $K(z)= \max\{K(x),K(y),K(z)\}$.
Further assume that $K(x) \geq K(y)$ (the remaining case is symmetrical).
Then, using  the symmetry of information to determine the maxima,
we also find $K(z \mid x^*) \geq K(x \mid z^*)$ 
and $K(z \mid y^*) \geq K(y \mid z^*)$. Then
the maxima in the terms of the equation $d(x,y) \leq d(x,z)+d(y,z)$ are
determined, and our proof obligation reduces to:
\begin{equation}\label{eq.req1}
\frac{K(x \mid y^*)}{K(x)}
\leq 
 \frac{K(z \mid x^*)}{K(z)} +
 \frac{K(z \mid y^*)}{K(z)},
\end{equation}
up to an additive term $O(1/K(z))$.
To prove (\ref{eq.req1}) we proceed as follows:

Applying the triangle inequality (\ref{eq.magic})  and dividing
both sides by $K(x)$, we have
\begin{equation}\label{eq.eq1}
\frac{K(x \mid y^*)}{K(x)}
\leq 
 \frac{K(x \mid z^*)+K(z \mid y^*)+O(1)}{K(x)}, 
\end{equation}
where the left-hand side is $\leq 1$. 

{\bf Case 2.1:} Assume that the right-hand side is $\leq 1$.
Setting $K(z) = K(x) + \Delta$, and  observe
$K(x|z^*)+\Delta = K(z|x^*)+O(1)$ 
by \eqref{eq.soi}. Add $\Delta$ to
both the numerator and the denominator 
in the right-hand side of \eqref{eq.eq1}, which increases the right-hand 
side because it is a ratio $\leq 1$, and rewrite:
\begin{eqnarray*}
\frac{K(x \mid y^*)}{K(x)}
&& \leq
 \frac{K(x \mid z^*) + K(z \mid y^*) + \Delta +O(1)}{K(x) + \Delta}
\\&& =
 \frac{K(z \mid x^*) + K(z \mid y^*)+O(1)}{K(z)},
\end{eqnarray*}
which was what we had to prove.

{\bf Case 2.2:} The right-hand side is $\geq 1$. We proceed like
in Case 2.1, and add $\Delta$ to both numerator and denominator.
Although now the right-hand side decreases, it must still
be $\geq 1$. 
This proves Case 2.2.
\end{proof}
\end{proof}

Clearly, $d(x,y)$ takes values in the range 
$[0,1+ O(1/\max\{K(x),K(y)\})]$.
To show that it is a normalized distance, it is left
to prove the density condition of Definition~\ref{eq.defsm}:

\begin{lemma}
The function $d(x,y)$ satisfies the density condition
\eqref{eq.dp}.
\end{lemma}
\begin{proof}
{\bf Case 1:}
Assume $K(y) \leq K(x)$.
Then, 
$d(x,y)=K(x \mid y^*)/K(x)$. 
If $d(x,y) \leq e$, then
$
K(x \mid y^*)  \leq  eK(x) .
$
Adding $K(y)$ to both sides, rewriting according to \eqref{eq.soi},
and subtracting $K(x)$ from both sides,
we obtain 
\begin{equation}\label{eq.case1}
K(y \mid x^*) \leq eK(x)+K(y)-K(x) \leq eK(x). 
\end{equation}
There are at most $\sum_{i=0}^{eK(x)} 2^i < 2^{eK(x)+1}$
binary programs of length $\leq eK(x)$. Therefore, for fixed $x$
there are $< 2^{eK(x)+1}$ objects $y$ satisfying \eqref{eq.case1}.

{\bf Case 2:}
Assume $K(x) < K(y)$.
Then,
$d(x,y)=K(y \mid x^*)/K(y)$.
If $d(x,y) \leq e$, then \eqref{eq.case1} holds again.
Together, Cases 1 and 2 prove the lemma.
\end{proof}

Since we have shown that $d(x,y)$ 
takes values in $[0,1]$, it satisfies the metric
requirements up to the given additive precision,
and it satisfies the density requirement in
Definition~\ref{eq.defsm}, it follows:
\begin{theorem}
The function $d(x,y)$ is a normalized distance
that satisfies the metric (in)equalities up to
$O(1/K)$ precision, where $K$ is the maximum of the
Kolmogorov complexities involved in the (in)equality
concerned.
\end{theorem}
\begin{remark}
\rm
As far as the
authors know, the idea of normalized metric is not
well-studied. An exception is \cite{Ya02}, which investigates
normalized metrics
to account for relative
distances rather than absolute ones, and it does so
 for much the same reasons as in
the present work. An example there is the normalized
Euclidean metric $|x-y|/(|x|+|y|)$, where $x,y \in {\cal R}^n$
(${\cal R}$ denotes the real numbers)  and $| \cdot |$ is the
Euclidean metric---the $L_2$ norm. Another example is a
normalized symmetric-set-difference metric.
But these normalized metrics are not necessarily effective in that
the distance between two objects gives the length of an effective
description to go from either object to the other one.
\end{remark}

\section{Universality}
We now show that $d(x,y)$ is universal then 
it incorporates every upper semi-computable (Definition~\ref{def.comp})
similarity in that
if objects $x,y$ are similar 
according to a particular
feature of the above type, then they are at least that
similar in the $d(x,y)$ sense.
We prove this by demonstrating that $d(x,y)$ is at least as small as any
normalized distance between 
$x,y$ in the wide class of upper semi-computable normalized
distances. This class is so wide that it will capture
everything that can be remotely of interest. 
\begin{remark}
\rm
The function $d(x,y)$ itself, being a ratio between two
maxima of pairs of upper semi-computable functions, may not itself
be semi-computable. (It is easy to see that this is likely, but
a formal proof is difficult.) In fact, $d(x,y)$ has ostensibly
only a weaker computability
property: Call a function $f(x,y)$
{\em computable in the limit} if there exists
a rational-valued recursive function $g(x,y,t)$ such that 
 $\lim_{t \rightarrow \infty} g(x,y,t)=f(x,y)$. 
Then $d(x,y)$ is in this class. It can be shown 
\cite{Ga01}
that this is precisely the class of functions
that are Turing-reducible
to the halting set.  While $d(x,y)$ is possibly not upper
semi-computable, it captures all similarities
represented by the upper semi-computable normalized distances 
in the class concerned, which
should suffice as a theoretical basis for all practical purposes.
\end{remark}


%
%
%
%
%
%
%
%
%
%
%

\begin{theorem}
The normalized information distance $d(x,y)$ 
minorizes every upper semi-computable
normalized distance 
$f(x,y)$ by
$d(x,y) \leq f(x,y) + O( 1/K )$ where 
$K= \min\{K(x),K(y)\}$.
\end{theorem}

\begin{proof}
Let $x,y$ be a pair of objects and let $f$ be a
normalized distance
that is upper semi-computable. Let $f(x,y)=e$. 

{\bf Case 1:} Assume that $K(x) \leq K(y)$.
Then, given $x$ we can
recursively enumerate the pairs $x,v$ such that 
$f(x,v) \leq e$.
Note that the enumeration contains $x,y$.
By the normalization condition \eqref{eq.dp}, 
the number of pairs enumerated
is less than $2^{eK(x)+1}$.
Every such pair, in particular $x,y$, 
can be described by its index of length 
$\leq eK(x)+1$ in this
enumeration. 
Since the Kolmogorov complexity is the length of the
shortest effective description, given $x$, 
the binary length of the index plus 
an $O(1)$ bit program to perform
the recovery of $y$, must at least be as 
large as the Kolmogorov complexity, which yields
$K(y \mid x)\leq  eK(x)+O(1)$. 
Since $K(x) \leq K(y)$, by \eqref{eq.soi}, $K(x \mid y^*) \leq K(y \mid x^*)$,
and hence $d(x,y)= K(y \mid x^*)/K(y)$. Note that 
$K(y \mid x^*) \leq K(y \mid x )+O(1)$, because $x^*$ supplies the information
$(x,K(x))$ which includes the information $x$. Substitution gives:
\begin{align*}
d(x,y) & = \frac{ K(y \mid x^*)}{K(y)} 
 \leq \frac{eK(x)+O(1)}{K(x)}
 \\& \leq f(x,y)  + O( 1/K(x)) .
\end{align*}

{\bf Case 2:} Assume that $K(x) > K(y)$. 
Then, given $y$ we can
recursively enumerate the pairs $u,y$ such that
$f(u,y) \leq e$.
Note that the enumeration contains $x,y$.
By the normalization condition \eqref{eq.dp},
the number of pairs enumerated
is less than $2^{eK(y)+1}$.
Every such pair, in particular $x,y$,
can be described by its index of length
$\leq eK(y)+1$ in this
enumeration.
Similarly to Case 1, this yields
$K(x \mid y)\leq  eK(y)+O(1)$. 
Also, by \eqref{eq.soi}, $K(y \mid x^*) \leq K(x \mid y^*)$,
and hence $d(x,y)= K(x \mid y^*)/K(x)$. Substitution gives:
\begin{align*}
d(x,y) & = \frac{ K(x \mid y^*)}{K(x)}
 \leq \frac{eK(y)+O(1)}{K(y)}
 \\& \leq f(x,y)  + O( 1/K(y)) .
\end{align*}
\end{proof}

\section{Application to Whole Mitochondrial Genome Phylogeny}

It is difficult to find a 
more appropriate type of object than DNA sequences
to test our theory: such sequences
are finite strings over a 4-letter alphabet that are naturally 
recoded as
binary strings with 2 bits per letter. 
We will use whole mitochondrial DNA genomes of 20 mammals and the
problem of Eutherian orders to experiment.
The problem we consider is this: 
It has been debated in biology which two of the three main groups of placental
mammals,
Primates, Ferungulates, and Rodents,
are more closely related.
One cause of debate is that the maximum likelihood method of phylogeny
reconstruction gives
(Ferungulates, (Primates, Rodents)) grouping for half of
the proteins in mitochondial genome, and 
(Rodents, (Ferungulates, Primates)) for the other half
\cite{Cao1998}.  The authors
aligned 12 concatenated    
mitochondrial proteins taken from the following species: rat ({\em Rattus
norvegicus}), house mouse ({\em Mus musculus}), grey seal ({\em
Halichoerus grypus}), harbor seal ({\em Phoca vitulina}), cat ({\em
Felis catus}), white rhino ({\em Ceratotherium simum}), horse ({\em
Equus caballus}), finback whale ({\em Balaenoptera physalus}), blue
whale ({\em Balaenoptera musculus}), cow ({\em Bos taurus}), gibbon
({\em Hylobates lar}), gorilla ({\em Gorilla gorilla}), human ({\em
Homo sapiens}), chimpanzee ({\em Pan troglodytes}), pygmy chimpanzee
({\em Pan paniscus}), orangutan ({\em Pongo pygmaeus}), Sumatran
orangutan ({\em Pongo pygmaeus abelii}), using opossum ({\em Didelphis
virginiana}), wallaroo ({\em Macropus robustus}) and platypus ({\em
Ornithorhynchus anatinus}) as the outgroup, and built the maximum
likelihood tree. The currently
accepted grouping is (Rodents, (Primates, Ferungulates)). 

\subsection{Alternative Approaches:}
Before applying our theory,
we first examine the alternative approaches, in addition to that
of \cite{Cao1998}.
The mitochondrial genomes of the above 20 species were obtained from
GenBank. Each is about 18k bases, and each base is one out of four types:
adenine (A), which pairs with thymine (T), 
and cytosine (C), which pairs with guanine (G).

\begin{figure}[t]
\hfill\ \psfig{figure=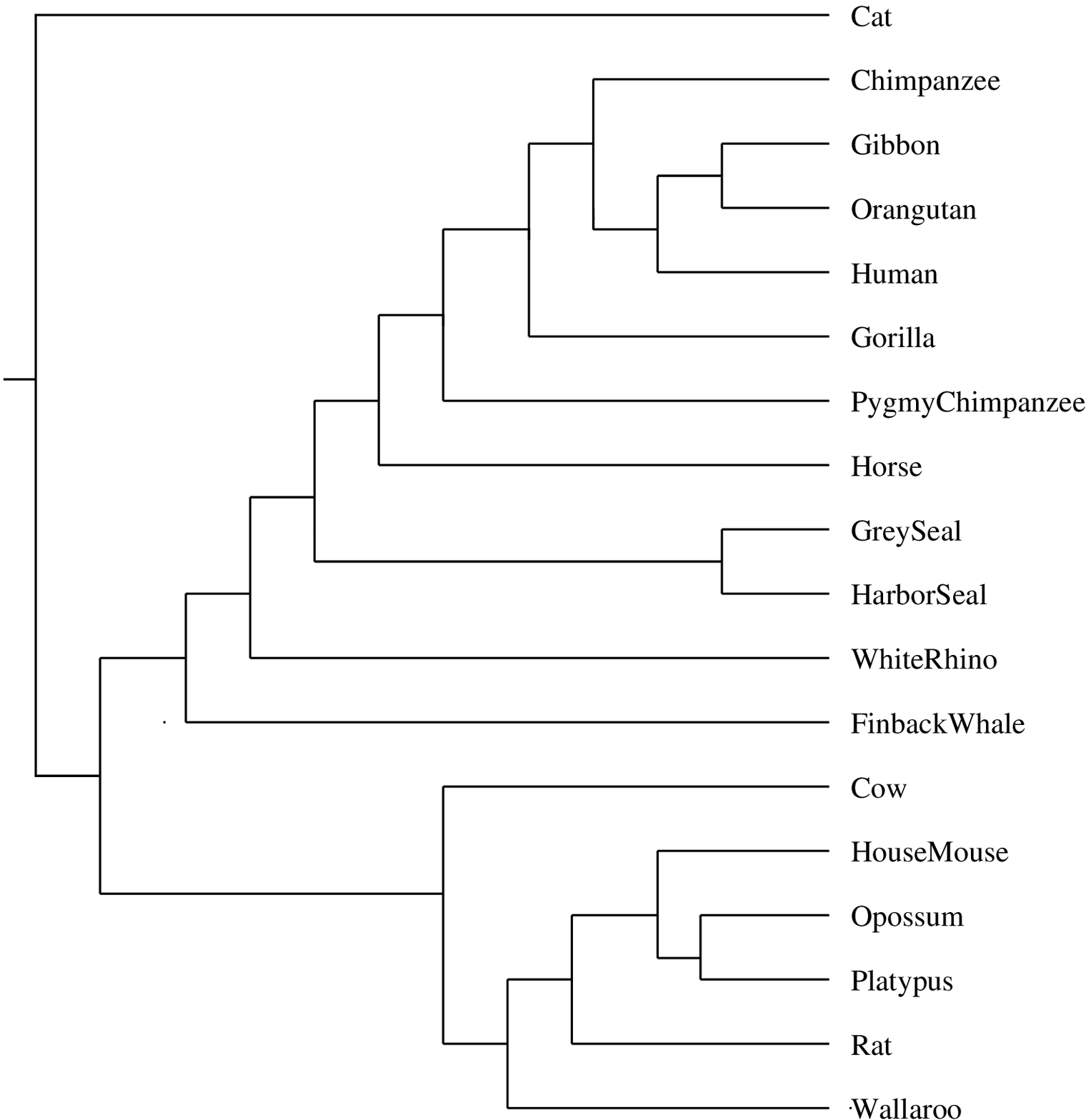,width=2.5in,height=2.5in} \hfill\
\caption{The evolutionary tree built from complete mammalian mtDNA
sequences using frequency of $k$-mers.}
\label{tree-frequency}
\end{figure}

{\bf k-mer Statistic:}
In the early years, researchers experimented using G+C contents,
or slightly more general $k$-mers (or Shannon block entropy)
to classify DNA sequences.
This approach uses the frequency statistics
of length $k$ substrings in a genome and the phylogeny is constructed
accordingly.
To re-examine this approach, we performed simple experiments: Consider
all length $k$ blocks in each mtDNA, for $k =1,2, \ldots , 10$.
There are $l=(4^{11}-1)/3$ different such blocks (some may not occur).
We computed the frequency of (overlapping) occurrences of each block
in each mtDNA. This way we
obtained a vector of length $l$ for each mtDNA, where the $i$th entry
is the frequency with which the $i$th block occurs overlapping
in the mtDNA concerned ($1 \leq i \leq l$). For two such vectors
(representing two mtDNAs) $p,q$, their distance is computed
as $d(p,q)=\sqrt { (p-q)^T (p-q) }$. Using neighbor joining 
\cite{Saitou1987}, the phylogeny tree that resulted is 
given in Figure \ref{tree-frequency}.
Using the hypercleaning method \cite{Bryant2000}, we obtain equally
absurd results. 
Similar experiments were repeated for size $k$ blocks alone 
(for $k=10,9,8,7,6$), without much improvement.

{\bf Gene Order:}
In \cite{Boore1998} the authors
propose to use the order of genes to infer the evolutionary 
history. This approach does not work for closely related species
such as our example where all genes are in the same order 
in the mitochondrial genomes in all 20 species.

{\bf Gene Content:}
The gene content method, proposed in \cite{Fitz,Snel1999},
uses as distance the ratio between the number of genes two species share 
and the
total number of genes. While this approach does not work here due to
the fact that all 20 mammalian mitochondrial genomes share exactly 
the same genes, notice the similarity of the gene content formula and our
general formula.

{\bf Rearrangement Distance:}
Reversal and rearrangement distances in
\cite{Kececioglu1995,Hannenhalli1995,Nadeau1998} compare genomes using
other partial genomic information such as the number of reversals
or translocations. These operations also do not appear in our
mammalian mitochondrial genomes, hence the method again is not
proper for our application.

{\bf Transformation Distance or Compression Distance:}
The transformation distance proposed in \cite{Varre1998} and
compression distance proposed in \cite{Grumbach1994} 
essentially correspond to $K(x|y)$ which is asymmetric, and so 
they are not admissible distances. Using $K(x|y)$ in the GenCompress 
approximation version produces a wrong tree
with one of the marsupials mixed up with ferungulates 
(the tree is not shown here).

\subsection{Our Compression Approach}
We have shown that the normalized information
distance $d$ (and up to a factor 2 this holds also for $d_s$)
is universal among the wide class normalized
distances, including all computable ones. 
These universal distances (actually, metrics) between $x$ and $y$
are expressed in terms
of $K(x),K(y)$, and $K(x \mid y)$.
The generality of the normalized information distance $d$ comes at the price
of noncomputability: 
Kolmogorov complexity is not computable but just upper semi-computable,
Section~\ref{sect.prel}, and $d$ itself is (likely to be) not even that.
Nonetheless, using standard compressors, we can compute
an approximation of $d$. 

\begin{remark}
\rm
To prevent confusion, we stress that, in principle, we cannot determine
how far a computable
approximation of $K(x)$ exceeds its true value. What we can say is that
if we flip a sequence $x$ of $n$ bits with a fair coin, then with overwhelming
probability we will have $K(x)$ is about $n$ and a real compressor
will also compress $x$ to a string of about length $n$ (that is, it
will not compress at all and the compressed file length is about
the Kolmogorov complexity and truely approximates it). However,
these strings essentially consist of random noise and have no meaning.
But if we take a meaningful string, for example the first $10^{23}$
bits of the binary representation of $\pi=3.1415 \ldots$, then
the Kolmogorov complexity is very short (because a program of,
say, 10,000 bits can compute the string), but no standard compressor
will be able to compress the string significantly below its length
of $10^{23}$
(it will not be able to figure out the inherent regularity).    
And it is precisely the rare meaningful strings, rare in comparison
to the overwhelming majority of strings that consist of random noise,
that we can be possibly interested in, and for which the Kolmogorov
complexity depends on computable regularities. Certain of those
regularities may be easy to determine, even by a simple compressor,
but some regularities may take
an infeasible amount of time to discover.
\end{remark}

It is clear how to compute the real-world compressor version
of the unconditional complexities involved. With
respect to the conditional complexities, by
$(\ref{eq.soi})$ we have $K(x \mid y) = K(x,y) - K(y)$
(up to an additive constant), and it is easy to see
that $K(x,y) = K(xy) $ up to additive logarithmic precision. (Here $K(xy)$
is the length of the shortest program to compute the concatenation
of $x$ and $y$ without telling which is which. To retrieve
$(x,y)$ we need to encode
the separator between the binary programs for $x$ and $y$.) 
So $K(x \mid y)$ is roughly equal to $K(xy)-K(y)$.

In applying
the approach in practice, we have to make do with an approximation based on a
real-world reference compressor $C$.
The resulting applied approximation of the ``normalized
information distance'' $d$ is called the
{\em normalized compression distance (NCD)}
\begin{equation}\label{eq.ncd}
NCD(x,y) = \frac{C(xy)- \min \{C(x),C(y)\}}{ \max\{C(x),C(y)\}}.
\end{equation}
Here,
$C(xy)$ denotes the compressed size of the concatenation of $x$ and $y$,
$C(x)$ denotes the compressed size of $x$,
and $C(y)$ denotes the compressed size of $y$.
The NCD is a non-negative number $0 \leq  r \leq 1 + \epsilon$ representing how
different the two files are. Smaller numbers represent more similar files.
The $\epsilon$ in the upper bound is due to
imperfections in our compression techniques,
but for most standard compression algorithms one is unlikely
to see an $\epsilon$ above 0.1 (in our experiments gzip and bzip2 achieved
NCD's above 1, but PPMZ always had NCD at most 1).

The theory as developed for the Kolmogorov-complexity based ``normalized
information distance'' in this paper
 does not hold directly for the (possibly poorly) approximating NCD.
In \cite{CV03}, we developed the theory of NCD based on the
notion of a ``normal compressor,'' and show that the NCD is a (quasi-)
universal similarity metric relative to a normal reference compressor $C$.
The NCD violates metricity only in sofar as it deviates from ``normality,''
and it violates universality only insofar as $C(x)$ stays above
$K(x)$.
The theory developed in the present paper is
the boundary case $C=K$, where the ``partially violated universality''
has become full ``universality''.
The conditional $C(y|x)$ has been replaced by $C(xy)-C(x)$,
which can be interpreted in stream-based compressors
as the compression length of $y$
based on using the ``dictionary'' extracted from $x$.
Similar statments hold for
block sorting compressors like bzip2, and 
designer compressors like GenCompress. Since the writing of this paper
the method has been released in the public domain as open-source software
at http://complearn.sourceforge.net/:
The CompLearn Toolkit is a suite
of simple utilities that one can use to apply compression
techniques to the process of discovering and learning patterns.
The compression-based approach used is powerful because it
can mine patterns in
in completely different domains.
In fact, this method is so general that it requires
no background knowledge about any particular
subject area. There are no domain-specific parameters to set,
and only a handful of general settings.

{\bf Number of Different k-mers:}
We have shown that using $k$-mer frequency statistics alone does not
work well. However, let us now combine the $k$-mer approach
with the incompressibility approach. 
Let the number
of distinct, possibly overlapping, $k$-length words in a sequence $x$ be $N(x)$.
With $k$ large enough, at least $\log_a(n)$, where $a$ is the
cardinality of the alphabet and $n$ the length of $x$,
we use $N(x)$ as a
rough approximation to $K(x)$. 
For example, for a sequence with the repetition of only one letter,
this $N(x)$ will be 1.
The length $k$ is chosen such that:
(i) If the  two genomes concerned would have been generated
randomly, then it is unlikely that they have a $k$-length word in common; and
(ii) It is usual that two homologous sequences share the same $k$-length words.
A good choice is $k= O(\log_4 n)$, where $n$ is the length of the genomes
and 4 is because we have 4 bases.
There are $4^{\log_4 n} = n$ subwords because
the alphabet has size 4 for DNA.  To describe a particular choice of
$N$ subwords of length $k=\log_4 n$ in a string of length $n$ we
need approximately $ \log {n \choose N} = N \log ( (4^k)/N ) = 2kN - N \log N$
bits.  
For a family of
mitochondrial DNA, we typically have
$5,000 \leq N,n  \leq 20,000$.  In this range, $2kN - N \log N$
can be approximated by $c N$ for some constant $c$.
So, overall the number of different subwords of length $k$ is 
proportional to $N$ for this choice of parameters.
                                                                                
According to our experiment, $k$ should be slightly larger than
$\log n$.  For example, a mitochondrial DNA is about 17K bases long.
$\log_4 17000 = 7.02$, while the $k$ we use  below is in range of
$6, \ldots, 13$, $7, \ldots,13$, or 
$8, \ldots, 13$, according to different formula and
whether spaced seeds (see below) are used.

\begin{figure}[t]
\hfill\ \psfig{figure=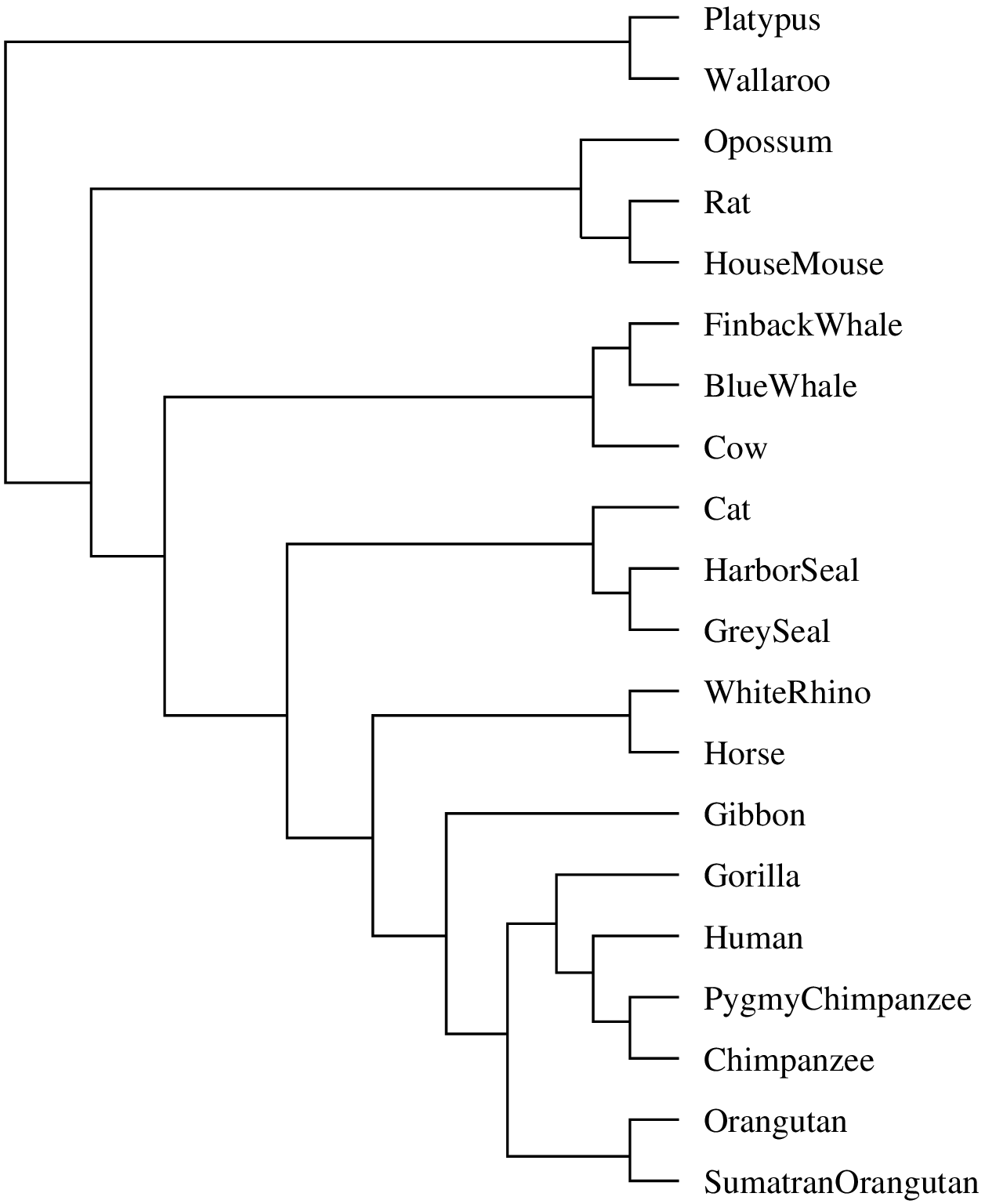,width=2.5in,height=2.5in} \hfill\
\caption{The evolutionary tree built from complete mammalian mtDNA
sequences using block size $k=7$ and $d'$.
}
\label{consmax7}
\end{figure}

\begin{figure}[t]
\hfill\ \psfig{figure=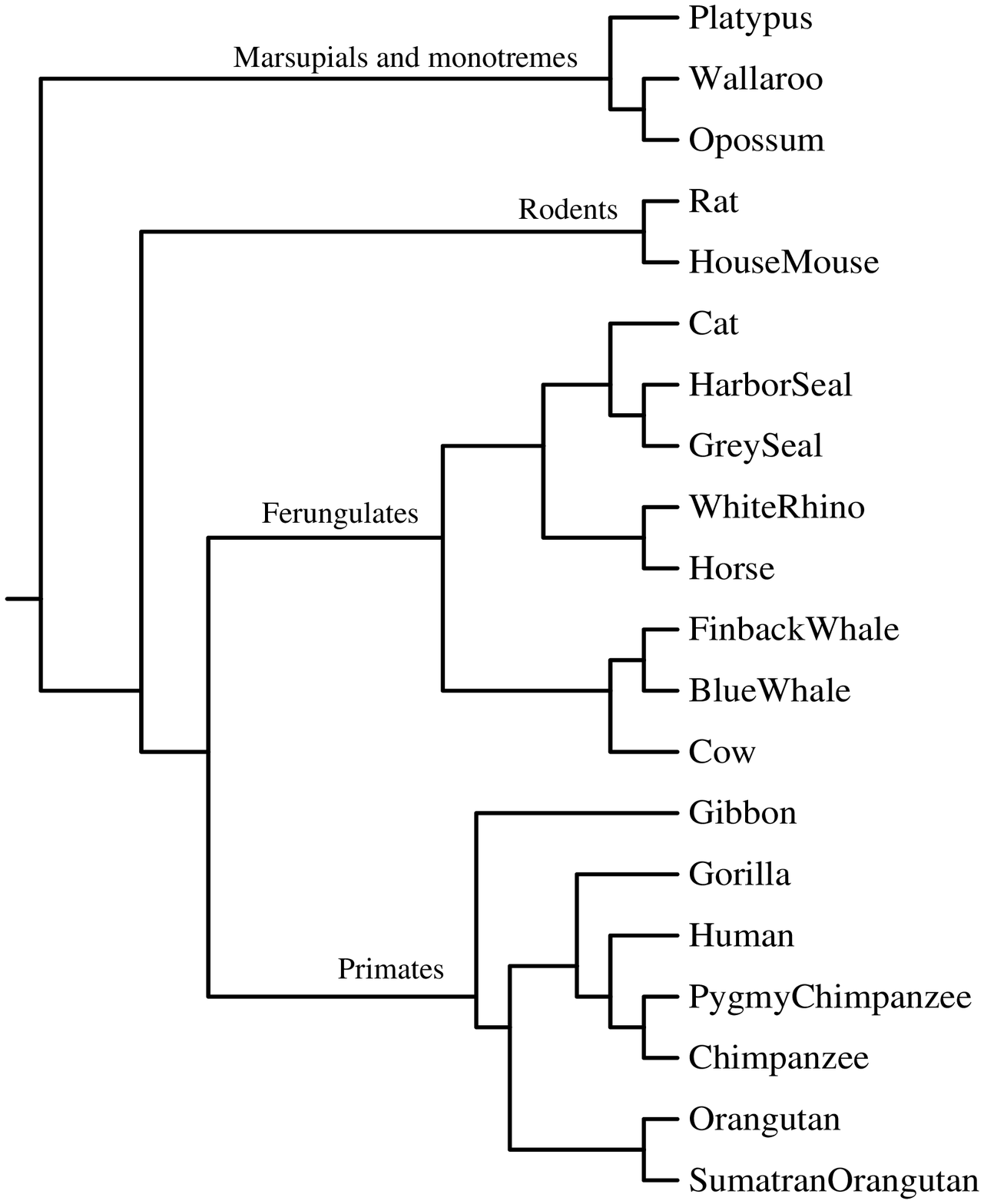,width=2.5in,height=2.5in} \hfill\
\caption{The evolutionary tree built from complete mammalian mtDNA
sequences. 
}
\label{tree-mammal}
\end{figure}

We justify the complexity approximation
using the number of different $k$-mers by the pragmatic observation 
that, because the genomes evolve by duplications, rearrangements and mutations,
\cite{Sa99},
and assuming that duplicated subwords are to be regarded as
duplicated information that can be ``compressed out,'' while distinct
subwords are not ``compressed out,''
it can be informally and intuitively argued 
that a description of the set of different subwords describes $x$.
With our choice of parameters 
it therefore is appropriate to use $N(x)$
as a plausible proportional estimate for $K(x)$ in case
$x$ is a genome.
So the size of the set is used to replace the $K(x)$ of genome $x$.
$K(x,y)$ is replaced by the size of the union of the two subword sets.
  Define $N(x|y)$ as $N(xy)-N(y)$.
Given two sequences $x$ and $y$, following the definition of  $d$,
(\ref{distance.m}),
the distance between $x$ and $y$ can be defined as
\begin{equation}\label{def.dprime}
d'(x,y) = \frac {\max \{ N(x|y), N(y|x)\}}{\max \{ N(x), N(y) \} }.
\end{equation}
Similarly, following $d_s$, (\ref{distance.s}) we can also
define another distance using $N(x)$,
\begin{equation}
d^*(x,y) = \frac { N(x|y) + N(y|x)} {N(xy)}.
\label{def-d*}
\end{equation}
Using $d'$ and $d^*$, we computed the distance matrixes for
the 20 mammal mitochondrial DNAs.  Then we used hyperCleaning
\cite{Bryant2000} to construct the phylogenies for the 20 mammals.
Using either of $d'$ and $d^*$, we were able to construct the tree 
correctly when $8\leq k \leq 13$, as in Figure~\ref{tree-mammal}.
A tree constructed with $d'$ for $k=7$ is given in 
Figure~\ref{consmax7}.  We note
that the opossum and a few other species are misplaced.  The tree 
constructed with $d^*$ for $k=7$ is very similar, but it correctly 
positioned the opossum.

{\bf Number of Spaced $k$-mers}
In methods for doing DNA homology search, a pair of 
identical words, each from a DNA sequence, is called a ``hit''.
Hits have been used as ``seeds'' to generate a longer match between
the two sequences.  If we define $N(x|y)$ as the number of distinct
words that are in $x$ and not in $y$, then the more hits the
two sequences have, the smaller the $N(x|y)$ and $N(y|x)$ are.
Therefore, the \eqref{def.dprime}, \eqref{def-d*} distances
can also be interpreted
as a function of the number of hits, each of which indicates some
mutual information of the two sequences.

As noticed by the authors of~\cite{MTL01}, though it is difficult 
to get the first hit (of $k$ consecutive letters)
in a region, it only requires one more base match 
to get a second hit overlapping the existing one.
This makes it inaccurate to attribute the same amount of information to 
each of the hits.  For this reason, we also tried to use the 
``spaced model'' introduced in~\cite{MTL01} to compute our distances.
A length-$L$, weight-$k$ spaced template is a 0-1 string of length $L$ 
having $k$ entries $1$.  We shift the template over the DNA sequence,
one position each step,
starting with the first positions aligned and finishing
with the last positions aligned. At each step extract the ordered
sequence of the $k$ bases
in the DNA sequence covered by the $1$-positions of the template
to form a length-$k$ word.  The number of different such
words is then used to define the distances $d'$ and $d^*$ in
Formula~(\ref{distance.s})~and~(\ref{def-d*}).

We applied the new defined distances to the 20 mammal data.  The performance
is slightly bettern than the
performance of the distances defined in (\ref{distance.s})~and~(\ref{def-d*}).
The modified $d'$ and $d^*$ can correctly construct the mammal tree when
$7\leq k \leq 13$ and $6\leq k \leq 13$, respectively.

{\bf Compression:}
To achieve the best approximation of Kolmogorov complexity,
and hence most confidence in the approximation of  $d_s$ and 
$d$, we used a new version of the {\em GenCompress} program,
\cite{Chen2000}, which achieved the best compression ratios 
for benchmark DNA sequences at the time of writing.
{\em GenCompress} finds approximate matches (hence edit distance becomes a
special case), approximate reverse complements, among other things,
with arithmetic encoding when necessary. 
Online service of {\em GenCompress} can be found on the web.
We computed $d (x,y)$ between each pair of
mtDNA $x$ and $y$, using {\em GenCompress} to heuristically
approximate $K(x|y)$, $K(x)$, and $K(x,y)$,
and constructed a tree (Figure~\ref{tree-mammal})
using the neighbor joining \cite{Saitou1987} program in the MOLPHY
package \cite{Adachi1996}. The tree is identical to the maximum
likelihood tree of Cao, {\it et al.} \cite{Cao1998}. For comparison,
we used the
hypercleaning program \cite{Bryant2000} and obtained the same result.
The phylogeny in Figure \ref{tree-mammal} re-confirms the hypothesis of
(Rodents, (Primates, Ferungulates)). Using the 
$d_s$ measure gives the same result.

To further assure our results, we have extracted only the coding regions 
from the mtDNAs of the above species, and performed the same computation.
This resulted in the same tree. 
\begin{remark}
\rm
In \cite{CV03} we have repeated these phylogeny experiments using bzip2 and PPMZ
compressors, and a new quartet method to reconstruct the phylogeny tree.
In all cases we obtained the correct tree. This is evidence that
the compression NCD method is robust under change of compressors, as long
as the window size of the used compressor is sufficient for the files
concerned, that is, GenCompress can be replaced by other more general-purpose
compressors. Simply use \cite{Ci03}.
\end{remark}

{\bf Evaluation:}
This new method for whole genome comparison and phylogeny
does not require gene identification nor any human intervention, in
fact, it is totally automatic. It is mathematically well-founded being
based on general information theoretic concepts. It works when there
are no agreed upon evolutionary models, as further demonstrated by the
successful construction of a chain letter phylogeny \cite{BLM03}
and when individual gene trees do not agree (Cao et al., 
\cite{Cao1998}) 
as is the case for genomes. As a next step, using the approach in 
\cite{CV03}, we have applied 
this method to much larger nuclear genomes of fungi and yeasts.
This work is not reported yet.

\section{The Language Tree}

\begin{figure}[t]\label{fig.tree1}
\hfill\ \psfig{figure=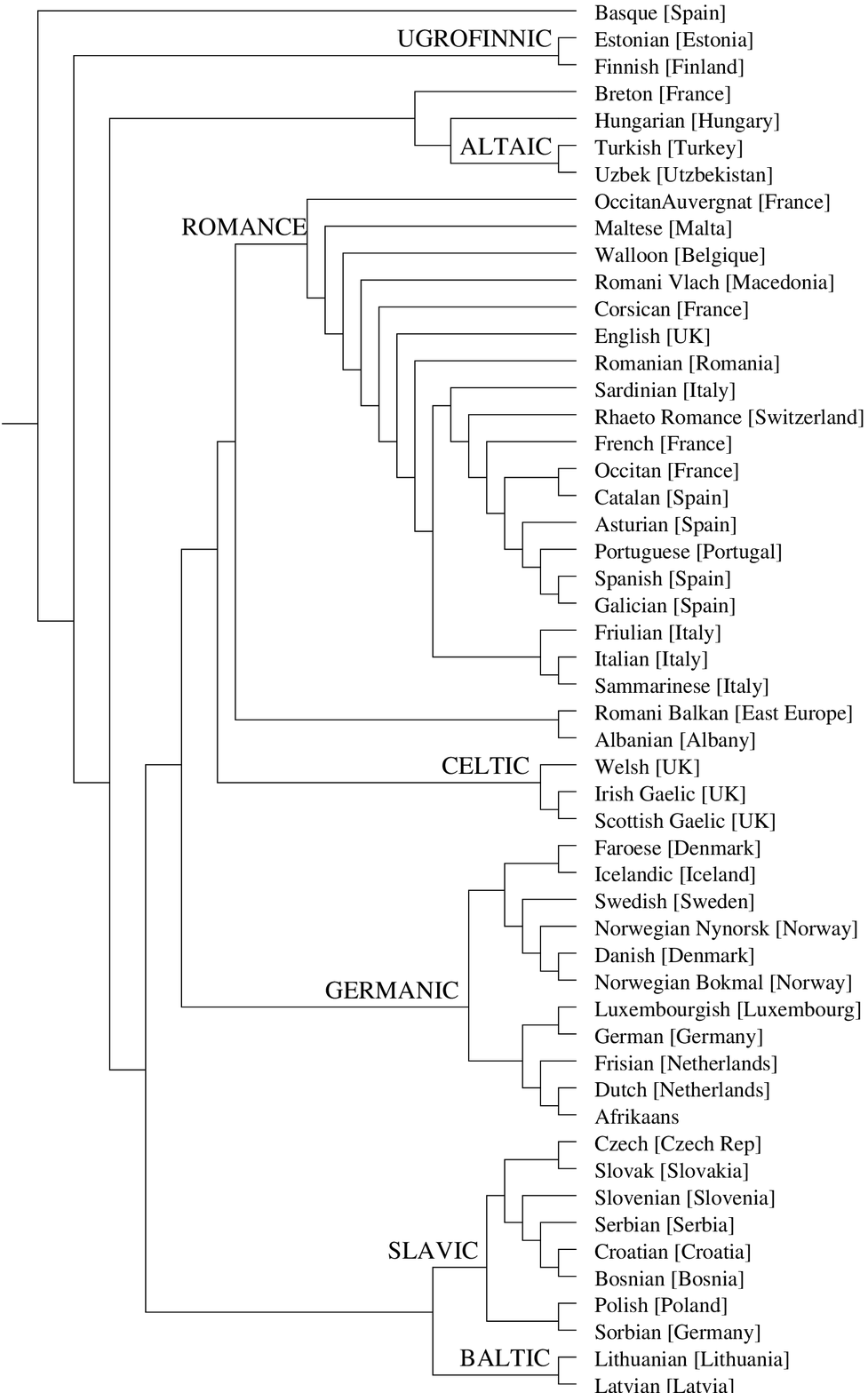,width=3in} \hfill\
\caption{The language tree 
using
approximated normalized information distance, $d_s$-version 
\protect\( \eqref{distance.s} \protect\), and neighbor 
joining.}
\end{figure}
\begin{figure}[h]\label{fig.tree2}
\hfill\ \psfig{figure=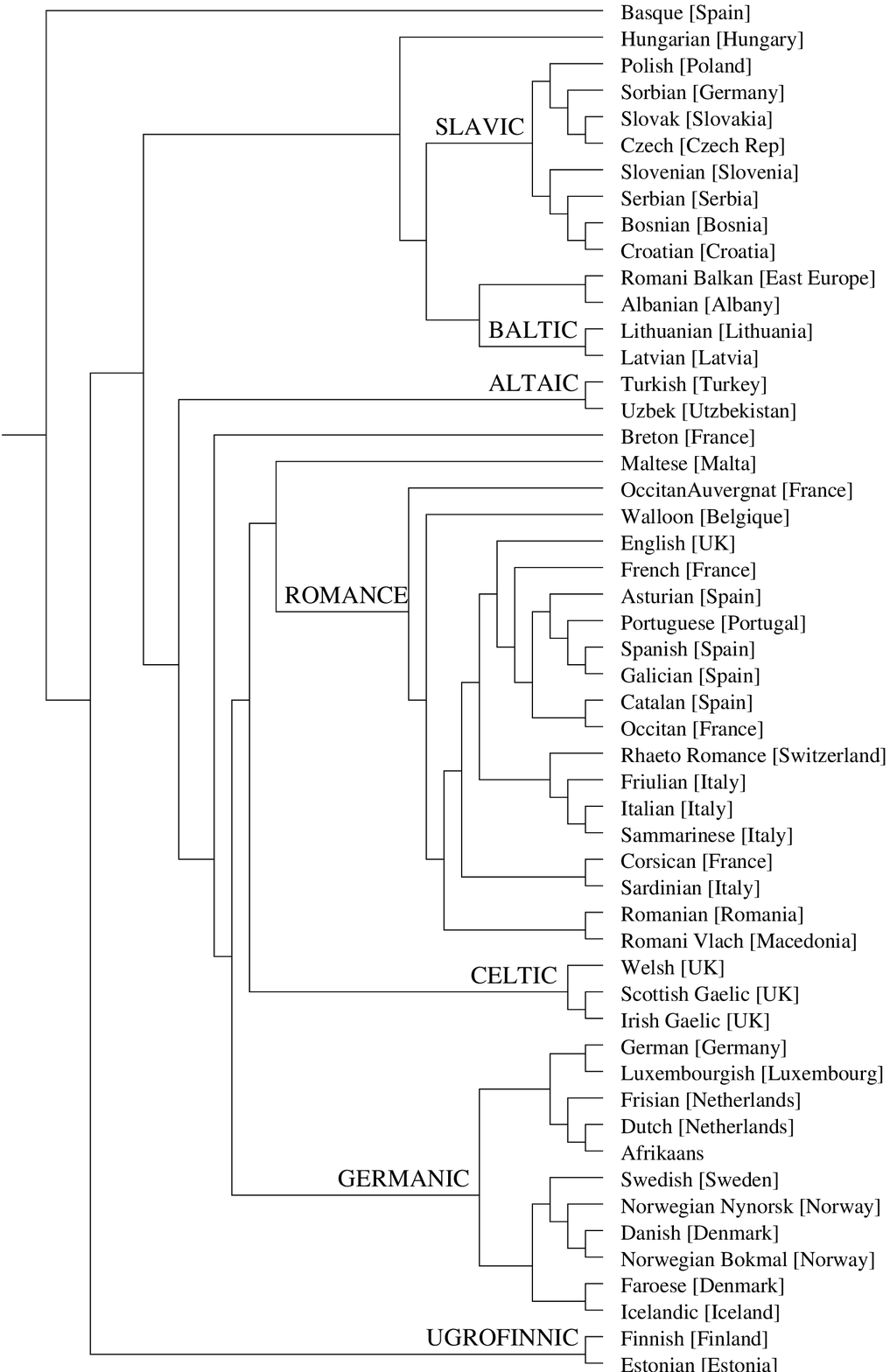,width=3in} \hfill\
\caption{The language tree 
using
approximated normalized information distance, $d$-version
\protect\( \eqref{distance.m} \protect\), and the
Fitch-Margoliash method.} 
\end{figure}
Normalized information distance is a totally general universal
tool, not restricted to a particular application area.
We show that it can also be used to successfully
classify natural languages. 
We downloaded the text corpora of  
``The
Universal Declaration of Human Rights'' in 52 Euro-Asian 
languages from the United Nations website \cite{rights}.
All of them are in UNICODE.
We first transform each UNICODE character
in the language text into an ASCII character by removing its
vowel flag if necessary. Secondly, as compressor to
compute the NCD we used a Lempel-Ziv compressor gzip.
This seems appropriate to compress these text corpora of
sizes (2 kilobytes) not exceeding the length of sliding window {\it gzip} uses
(32 kilobytes).
In the last
step,  we applied the $d_s$-metric
(\ref{distance.s}) with the neighbor-joining
package to obtain Figure~\ref{fig.tree1}. Even better worked
applying the $d$-metric 
(\ref{distance.m}) with the 
Fitch-Margoliash method \cite{FM1967} in the package PHYLIP
\cite{Adachi1996}); the resulting language classification tree
is given in Figure \ref{fig.tree2}. We note that all the main
linguistic groups can be successfully recognized, which includes
Romance, Celtic, Germanic, Ugro-Finnic, Slavic, Baltic, Altaic as
labeled in the figure.
In both cases, it is a
rooted tree using Basque [Spain] as outgroup.
The branch lengths are not proportional to the
actual distances in the distance matrix.

Any language tree built by only analyzing contemporary natural text corpora 
is partially corrupted by historical inter-language contaminations. In fact,
this is also the case with genomic evolution:
According to current insights phylogenetic trees are not only based
on inheritance, but also the environment is at work through selection,
and this even introduces an indirect interation between species, 
called reticulation\footnote{Joining of separate lineages on a 
phylogenetic tree, generally through hybridization or through 
lateral gene transfer. Fairly common in certain land plant clades; 
reticulation is thought to be rare among metazoans.\cite{Be04}} 
(arguably less direct than de borrowings
between languages).
Thus, while English is
ostensibly a Germanic Anglo-Saxon language, 
it has absorbed a great deal of French-Latin components. Similarly,
Hungarian, often considered a Finn-Ugric language, which consensus
currently happens to be open to debate in the linguistic community,
is known to have absorbed many Turkish and
Slavic components. Thus, an automatic construction of a language tree
based on contemporary text corpora,
exhibits current linguistic relations 
 which do not
necessarily coincide completely with the  historic language family
tree. 
The misclassification
of English as Romance language 
is reenforced by
the fact that the English vocabulary in the 
Universal Declaration of Human Rights, being nonbasic
in large part, is Latinate in large part.  This presumably also 
accounts for the misclassification of 
Maltese, an Arabic dialect with lots of Italian loan words,
as Romance.
Having voiced these caveats, the result of our automatic
experiment in language tree reconstruction is accurate.

Our method  improves the results of
\cite{BCL02a},
 using the same linguistic corpus,
using an asymmetric measure 
based on the approach sketched in the section ``Related Work.''
In the resulting language tree, English 
is isolated between Romance and Celtic languages, 
Romani-balkan and Albanian are isolated, and
Hungarian is grouped with Turkish and Uzbek.
The (rooted) trees resulting from our experiments (using Basque
as out-group) seem
more correct. We use Basque as outgroup since linguists regard
it as a language unconnected to other languages.

\section{Conclusion}
We developed a mathematical theory of compression-based similarity distances
and shown that there is a universal similarity metric:
the normalized information distance.  This distance uncovers all
upper semi-computable similarities, and therefore
estimates an evolutionary or relation-wise distance on strings. 
A practical version was exhibited based on standard compressors.
Here it has been shown
to be applicable to 
whole genomes, 
and to built a large language family tree from
text corpora. References to applications in a plethora of other fields
can be found in the Introduction.
It is perhaps useful to point out that
the results reported in the figures were obtained 
at the very first runs and have not been selected by appropriateness
from several trials. From the theory point-of-view
we have obtained a general mathematical theory forming a solid framework
spawning practical tools applicable in many fields. 
Based on the noncomputable notion
of Kolmogorov complexity, the normalized information distance
can only be approximated without
convergence guarantees. Even so,
the fundamental rightness of the approach is evidenced by the remarkable
success (agreement with known phylogeny in biology)
of the evolutionary trees obtained and the building of 
language  trees. 
From the applied side of genomics our work gives the first fully automatic
generation of whole genome mitochondrial phylogeny; in computational
linguistics it presents a fully automatic way to build language trees
and determine language families.

\appendix
\section{A Variant Method in Linguistics}
\label{app.A}
In \cite{BCL02a} 
the purpose is to infer a language tree from different-language text corpora,
as well as do 
authorship attribution on basis of text corpora.
The distances determined between objects are
justified by ad-hoc plausibility arguments
(although the information distance of \cite{LiVi97,BGLVZ98} 
is also mentioned).
The paper \cite{BCL02a} is predated by our universal similarity metric
work and phylogeny tree (hierarchical clustering) 
experiments \cite{CKL99,Chen2000,LV01}, but 
it is the language tree experiment we repeated in the present 
paper using our  
own technique with somewhat better results.
For comparison of the methods
we give some brief details.
Assume a fixed compressor 
(\cite{BCL02a,BCL02b} use the Lempel-Ziv type). Let
$C(x)$ denote the length of of the compressed version of a file $x$,
and let $x'$ be a short file from the same source as $x$. For example
if $x$ is a long text in a language, then $x'$ is a short text in the
same language. (The authors refer to sequences generated 
by the same ergodic source.)
Then two distances are considered between files $x,y$: 
(i) the asymmetric distance
$s(x,y)= ([C(xy')-C(x)] - [C(yy')-C(y)])/|y'|$, the numerator
quantifying the difference in compressing $y'$ using a data base
sequence generated by a different source versus one
generated by the same source that generated $y'$;
and a symmetric distance (ii) $S(x,y)= s(x,y) |y'|/[C(yy')-C(y)] +
s(y,x) |x'|/[C(xx')-C(x)]$. The distances are not metric
(neither satisfies the triangular inequality) and the authors
propose to ``triangularize'' in practice by a Procrustes method: setting
$S(x,y) := \min_w (S(x,w)+S(w,y))$ in case the left-hand side exceeds
the right-hand side. We remark that in that case the left-hand side 
$S(x,y)$ becomes smaller and may
in its turn cause a violation of another triangular inequality
as a member of the right-hand side, and so on.
On the upside, despite the lack of supporting theory,
the authors report successful experiments.

\section{A Variant Method in Data Mining}
\label{app.B}
In the follow-up data mining paper \cite{KLR04} the authors report successful
experiments using a simplified version of the NCD \eqref{eq.ncd} called
compression-based dissimilarity measure (CDM):
\[
CDM(x,y)= \frac{C(xy)}{C(x)+C(y)}.
\]
Note that this measure always ranges between $\frac{1}{2}$
(for $x=y$) and $1$ (for $x$ and $y$ satisfy $C(xy)=C(x)+C(y)$,
that is, compressing $x$ doesn't help in compressing $y$).
The authors don't give a theoretical analysis, but intuitively
this formula measures similarity of $x$ and $y$ by comparing 
the lengths of the compressed files in combination and seperately. 
 
{\small
\section*{Acknowledgement}
John Tromp carefully read and commented on an early draft,
and Teemu Roos supplied reference \cite{Ra61}.

\section{Biographies of the Authors}

{\sc Ming Li} is a CRC Chair Professor in Bioinformatics, of Computer Science
at the University of Waterloo. He is a recipient of
Canada's E.W.R. Steacie Followship Award in 1996, and the 2001 Killam
Fellowship. Together with Paul Vitanyi they pioneered applications of
Kolmogorov complexity and co-authored the book "An Introduction to
Kolmogorov
Complexity and Its Applications" (Springer-Verlag, 1993, 2nd Edition, 1997).
He is a co-managing editor of Journal of Bioinformatics
and Computatational Biology. He currently also serves on the editorial
boards of Journal of Computer and System Sciences, Information
and Computation, SIAM Journal on Computing, Journal of Combinatorial
Optimization, Journal of Software, and Journal
of Computer Science and Technology.

{\sc Xin Chen} received his Ph.d. from Peking University, Beijing, China, in
2001. He is now a Post-doc at University of California, Riverside. His
research interests include data compression, pattern recognition and
bioinformatics.

{\sc Xin Li} 
obtained  his B.Sc. degree in Computer Science from McMaster
University (Canada) and his M.Sc. degree in Computer Science from the
University of Western Ontario (Canada).

{\sc Bin Ma} received his Ph.D. degree from Peking University 
in 1999, and has been
an assistant professor in the Department of Computer
Science at the University of Western Ontario since 2000. He
is a recipient of Ontario
Premier's Research Excellence award in 2003 for his research in
bioinformatics.  He is a coauthor of two well-known bioinformatics
software programs, PatternHunter and PEAKS.

{\sc Paul M.B. Vit\'anyi} is a Fellow
of the Center for Mathematics and Computer Science (CWI)
in Amsterdam and is Professor of Computer Science
at the University of Amsterdam.  He serves on the editorial boards
of Distributed Computing (until 2003), Information Processing Letters, 
Theory of Computing Systems, Parallel Processing Letters,
International journal of Foundations of Computer Science,
Journal of Computer and Systems Sciences (guest editor),
and elsewhere. He has worked on cellular automata, 
computational complexity, distributed and parallel computing, 
machine learning and prediction, physics of computation,
Kolmogorov complexity, quantum computing. Together with Ming Li 
they pioneered applications of Kolmogorov complexity 
and co-authored ``An Introduction to Kolmogorov Complexity 
and its Applications,'' Springer-Verlag, New York, 1993 (2nd Edition 1997),
parts of which have been translated into Chinese,  Russian and Japanese.

\end{document}